\documentclass[a4paper,11pt]{article}
\usepackage{jcappub} 
\usepackage{lineno}
\usepackage{multirow}
\usepackage{array}
\usepackage{graphics}

\makeatletter
\input{aas_macros.sty}
\let\jnl@style=\relax
\makeatother

\title{HOD-informed prior for EFT-based full-shape analyses of LSS}

\author[a,b]{Hanyu Zhang,}
\author[a,b]{Marco Bonici,}
\author[c,d]{Guido D'Amico,}
\author[e]{Simone Paradiso,}
\author[a,b,f]{Will J. Percival}
\affiliation[a]{Waterloo Centre for Astrophysics, University of Waterloo,\\200 University Ave W, Waterloo, ON N2L 3G1, Canada}
\affiliation[b]{Department of Physics and Astronomy, University of Waterloo,\\200 University Ave W, Waterloo, ON N2L 3G1, Canada}
\affiliation[c]{Department of Mathematical, Physical and Computer Sciences, University of Parma,\\ Parco Area delle Scienze 7/a, 43124 Parma, Italy}
\affiliation[d]{INFN Gruppo Collegato di Parma,\\Parco Area delle Scienze 7/a, 43124 Parma, Italy}
\affiliation[e]{INAF, Istituto di Astrofisica Spaziale e Fisica Cosmica di Bologna,\\via P. Gobetti 101, I-40129 Bologna, Italy}
\affiliation[f]{Perimeter Institute for Theoretical Physics,\\31 Caroline St. North, Waterloo, ON N2L 2Y5, Canada}

\emailAdd{hanyu.zhang@uwaterloo.ca}

\abstract{To improve the performance of full-shape analyses of large-scale structure, we consider using a halo occupation distribution (HOD)-informed prior for the effective field theory (EFT) nuisance parameters. We generate 320\,000 mock galaxy catalogs using 10\,000 sets of HOD parameters across 32 simulation boxes with different cosmologies. We measure and fit the redshift-space power spectra using a fast emulator of the EFT model, and the resulting best-fit EFT parameter distributions are used to create the prior. This prior effectively constrains the EFT nuisance parameter space, limiting it to the space of HOD-mocks that can be well fit by a EFT model. We have tested the stability of the prior under different configurations, including the effect of varying the HOD sample distribution and the inclusion of the hexadecapole moment. We find that our HOD-informed prior and the cosmological parameter constraints derived using it are robust. While cosmological fits using the standard EFT prior suffer from prior effects, sometimes failing to recover the true cosmology within Bayesian credible intervals, the HOD-informed prior mitigates these issues and significantly improves cosmological parameter recovery for $\Lambda$CDM and beyond. This work lays the foundation for better full-shape large-scale structure analyses in current and upcoming galaxy surveys, making it a valuable tool for addressing key questions in cosmology.}

\begin{document}
\maketitle
\flushbottom

\section{Introduction}
\label{sec:intro}

The large-scale structure (LSS) of the universe, characterized by the distribution of galaxies and matter over vast cosmic distances, offers a profound window into the underlying physics governing the universe. By studying the LSS, cosmologists can glean insights into the nature of dark matter, dark energy, and the initial conditions of the universe. Previous-generation galaxy surveys, such as the Baryon Oscillation Spectroscopic Survey (BOSS) \cite{Dawson2013TheSDSS-III} and the extended BOSS (eBOSS) \cite{Dawson2016TheData}, have already made significant contributions to our understanding of cosmic expansion and structure growth. Building on these recent advancements, are the ongoing Dark Energy Spectroscopic Instrument (DESI) \cite{Levi2013The2013,DESICollaboration2016TheDesign,DESICollaboration2016TheDesignb,Collaboration2022OverviewInstrument} and Euclid mission \cite{EuclidCollaboration2024Euclid.Mission}, and the planned Rubin Observatory Legacy Survey of Space and Time (LSST) \cite{Ivezic2019LSST:Products}, and Nancy Grace Roman Space Telescope \cite{Spergel2015Wide-FieldReport}. These surveys aim to map the distribution of millions of galaxies with unprecedented precision, providing information about the distribution of perturbations in the early universe from the late-time clustering, and information about the late time universe from its projection into observed angular positions and redshifts of galaxies.

To analyze the wealth of data from galaxy surveys, the effective field theory of large-scale structure (EFTofLSS) \cite{Baumann2012CosmologicalFluid,Carrasco2012TheStructures,Porto2014TheStructures,Perko2016BiasedStructure} has emerged as a robust framework. EFTofLSS systematically incorporates non-linear gravitational effects and other complex phenomena that influence the formation and distribution of galaxies. This theoretical approach has proven instrumental in translating observed galaxy clustering patterns into constraints on cosmological parameters, as demonstrated in analyses of BOSS and eBOSS data \cite{Colas2020EfficientStructure,dAmico2020TheStructure,Ivanov2020CosmologicalSpectrum,DAmico2021LimitsCode,Niedermann2021NewData,Kumar2022UpdatingGalaxies, Simon2022ConstrainingStructures,Nunes2022NewSpectrum,Philcox2022BOSSMonopole,Zhang2022BOSSStructure,Chen2022ABAO,Lague2022ConstrainingSurveys,Simon2023ConsistencySpectrum,Carrilho2023CosmologyPriors,Schoneberg2023ComparativeRadiation,Smith2023AssessingConstant,Allali2023DarkDatasets,Simon2023CosmologicalAnalysis,Simon2023UpdatedEnergy,DAmico2024LimitsEFTofLSS}. Within the framework of $\Lambda$CDM, these analyses have provided strong constraints on the matter density parameter $\Omega_m$, comparable to Planck \cite{Collaboration2020PlanckParameters}, and on the Hubble constant $H_0$, with precision similar to that of the SH0ES collaboration \cite{Riess2019LargeCDM}.

While the EFTofLSS offers a state-of-the-art description of the physics of large-scale structure tracers, it necessitates an extensive number of nuisance parameters that need to be accounted for. These parameters, which encapsulate various small-scale effects, have sufficient effect on the model that they can lead to an estimate of cosmological parameters which is biased with respect to the truth if one considers the 1D Bayesian credible intervals if not properly constrained \cite{Simon2023ConsistencySpectrum,Gomez-Valent2022FastCosmology,Carrilho2023CosmologyPriors,DAmico:2022osl}. These issues, collectively referred to as prior effects \cite{Simon2023ConsistencySpectrum}, include the projection effect, where a large parameter space introduces degeneracies that affect the determination of the cosmological parameters, and the prior weight effect, where the weighting of different regions of the parameter space biases the posterior distribution. \cite{Holm2023BayesianData} showed that Bayesian and frequentist approaches can lead to different cosmological constraints due to the sensitivity of the Bayesian method to prior effects, resulting in significant differences between the credible and confidence intervals that result from fits using the two statistical methods.

To address this challenge, it is essential to constrain the nuisance parameter space effectively, reducing the risk of prior effects. One alternative approach to modeling the non-linear clustering is to use the HOD model \cite{Jing1998SpatialSurvey,Peacock2000HaloBias,Seljak2000AnalyticClustering,Scoccimarro2001HowClustering,Cooray2002HaloStructure,Berlind2002TheMass,Zheng2005TheoreticalGalaxies,Zheng2007GalaxyClustering,Zheng2009HaloGalaxies}, in which galaxies are placed in dark matter halos based on probabilistic prescriptions that depend on the properties of the host halo and its neighborhood. The HOD method has been successfully used to describe the small-scale non-linearity of biased tracers in both DESI and (e)BOSS \cite{White2011TheData, Richardson2012TheQuasars, Zhai2017TheData, Alam2020MultitracerEBOSS, Avila2020TheGalaxies, Smith2020TheSample, Zhou2021TheRedshifts, Rossi2021TheMeasurements, Zhang2022ConstrainingStatistics, Yuan2024TheABACUSSUMMIT,Rocher2023TheSimulations}. By leveraging the HOD model, we can generate mock galaxy catalogs that span a broad range of possible galaxy distributions, thereby providing a distribution of possible clustering scenarios that all have a physical basis, assuming HOD is a reasonable parametric description of galaxy formation up to small scales.

In this paper, we use a sample of such scenarios to derive an HOD-informed prior for the EFTofLSS model to address the issue of an overly large nuisance parameter volume. Our method involves generating a diverse set of galaxy power spectrum measurements from HOD mock catalogs, covering a wide range of HOD parameter space and various cosmological models. By fitting the EFTofLSS model to the power spectra derived from these mock catalogs, we identify the best-fit EFTofLSS parameters and use their distribution to form our HOD-informed prior \cite{2021JCAP...11..026S}. This approach ensures that the HOD-informed prior encompasses most of the physically plausible power spectra, allowing us to delineate a more constrained and physical parameter space. Regions outside this prior are likely unphysical and can be excluded from the analysis, thereby helping to reduce the prior effect and improve the precision of cosmological parameter estimates. We carefully tested several configurations that could influence the prior, including the impact of different cosmologies, variations in HOD parameter samples, the inclusion or exclusion of the hexadecapole moment, and dependence on covariance matrices. These tests ensure the robustness and reliability of the HOD-informed prior across various scenarios. 

In \cite{Ivanov2024Full-shapeBOSS,Ivanov2024Full-shapeAnomaly}, a similar strategy was adopted by incorporating simulation-based priors into their full-shape analysis, calibrated via a field-level forward model. This approach reduced the prior volume for bias parameters and led to substantial improvements in constraints on $\Omega_m$ and $\sigma_8$. They also explored the cosmology dependence of their learned prior and found it to be mild. Our work leverages a novel EFTofLSS emulator, allowing us to stack fits from a broad range of cosmological models, thereby covering a wider physical volume and making our priors less sensitive to a cosmological baseline.

Applying our HOD-informed prior to fit a mock catalog of luminous red galaxies (LRG), we observed a significant improvement in both accuracy and precision for the recovered cosmological parameters compared to the standard EFT prior for full-shape analysis. In particular, for the $\Lambda$CDM model, the prior effect was notably reduced. For instance, the standard prior resulted in a 1.65$\sigma$ deviation between the marginalized mean and the true value for $\Omega_c h^2$, while the HOD-informed prior reduced this to 0.57$\sigma$. Similarly, for $H_0$, the deviation was reduced from 1.55$\sigma$ with the standard prior to 0.87$\sigma$ with the HOD-informed prior. In the $w$CDM model, which includes additional parameter space, the prior effects were more pronounced. The standard prior resulted in deviations of 1.90$\sigma$ for $\Omega_c h^2$ and 2.55$\sigma$ for $\ln 10^{10} A_s$, whereas the HOD-informed prior reduced these deviations to 1.03$\sigma$ and 1.55$\sigma$, respectively, showing better performance in mitigating prior effects in this expanded parameter space. For the more complex $w_0 w_a$CDM model, while the data does not provide sufficient constraining power, our HOD-informed prior still performed better than the standard prior, consistent with the improvements observed in the $\Lambda$CDM and $w$CDM cases.

The structure of this paper is as follows: Section~\ref{sec:theory} provides a detailed overview of the theoretical framework, including the HOD model and the EFTofLSS. Section~\ref{sec:method} outlines our methodology, describing the generation of galaxy power spectra from HOD and N-body simulations, the rapid best-fit estimation of EFT parameters, and the process of learning the distribution with normalizing flow. Section~\ref{sec:prior} analyzes the impact of different configurations on the HOD-informed prior, including variations in HOD sample distributions, cosmological dependencies, and the inclusion or exclusion of the hexadecapole moment. Section~\ref{sec:comparison} presents a comparative analysis of priors on mock datasets, and Section~\ref{sec:conclusion} concludes with a summary of our findings and their implications for future full-shape LSS analysis.

\section{Theoretical Framework}
\label{sec:theory}

In this section, we introduce the models and theories that underpin our analysis. We begin with the galaxy-halo connection, outlining the HOD approach, which describes the relationship between galaxies and their underlying dark matter halos. This model is essential for generating realistic galaxy power spectrum measurements from dark matter based simulations. Following this, we briefly summarize the EFTofLSS, which provides a robust framework for modeling the clustering of galaxies on quasi-linear scales, incorporating various physical effects that influence the power spectrum. 

\subsection{HOD model}
\label{subsec:hod}

The HOD model is a statistical framework used to describe how galaxies occupy dark matter halos. By specifying the probability distribution of galaxy numbers within halos based on a set of given halo properties, the HOD provide a link between the distribution of dark matter and the observable clustering of galaxies. 

In this analysis, we adopt the HOD prescription described in \cite{Yuan2024TheABACUSSUMMIT} for LRGs. Building on the basic model originally proposed by \cite{Zheng2007GalaxyClustering}, which assumes that the expected number of galaxies in a dark matter halo depends solely on the halo mass, we also introduce flexibility in modeling galaxy velocities through velocity bias. Incorporating velocity bias is essential for generating the redshift-space clustering, which accounts for redshift-space distortions (RSD). The HOD model can be summarized as follows:
\begin{align}
    \langle N_{\mathrm{cen}}\rangle(M) &= \frac{1}{2} \mathrm{erfc} \left(\frac{\log_\mathrm{10}(M_{\rm cut}/M)}{\sqrt{2}\sigma}\right)\,, \label{eqn:Ncent}\\
    \langle N_{\mathrm{sat}}\rangle(M) &= \langle N_{\mathrm{cen}}\rangle(M)\left(\frac{M - \kappa M_{\rm cut}}{M_1}\right)^\alpha\,. \label{eqn:Nsate}
\end{align}
The expectation value differs for central galaxies, which occupy the center of the halo, and satellite galaxies, which are in motion around the center. Here, $M$ represents the mass of the host halo, and $M_\mathrm{cut}$ is the characteristic minimum mass required to host a galaxy. The parameter $\sigma$ describes the steepness of the probability increases with halo mass around $M_\mathrm{cut}$. $\kappa M_{\rm cut}$ sets a mass threshold for the satellite galaxies, while $\alpha$ controls how steeply the probability of hosting satellites rises with increasing halo mass. $M_1$ represents the extra mass above the threshold required for a halo to host, on average, one satellite galaxy. 

Regarding velocity bias, we focus on modifying the velocity along the line-of-sight (LoS) direction using two additional parameters,
\begin{align}
    v_{\mathrm{cen,LoS}}^{b} &= v_{\mathrm{halo,LoS}} + \alpha_{\mathrm{cen}} \delta v (\sigma_{\mathrm{LoS}})\,, \label{eqn:vcen}\\
    v_{\mathrm{sat,LoS}}^{b} &= v_{\mathrm{halo,LoS}} + \alpha_{\mathrm{sat}} (v_{\mathrm{sat,LoS}}-v_{\mathrm{halo,LoS}})\,. \label{eqn:vsat}
\end{align}
LoS in a simulation box is typically assumed to be a fixed, parallel direction, representing the path along which an observer at an infinite distance would view the system. In eq.~\ref{eqn:vcen} and \ref{eqn:vsat}, $v_{\mathrm{halo,LoS}}$ denotes the LoS component of the host halo's velocity, while $\delta v (\sigma_{\mathrm{LoS}})$ represents the velocity dispersion of the host halo. The term $v_{\mathrm{sat,LoS}}$ refers to the LoS component of the satellite galaxy's velocity before applying velocity biasing. It is important to note that different simulations may employ various strategies for velocity assignment. In general, the velocity bias parameters $\alpha_{\mathrm{cen}}$ and $\alpha_{\mathrm{sat}}$ describe deviations in velocity for central galaxies from the halo center and for satellite galaxies from the local dark matter environment, respectively.

As presented, the HOD model chosen for this analysis includes five HOD parameters adopted from \cite{Zheng2007GalaxyClustering}: $M_{\mathrm {cut}},M_1,\sigma,\alpha$,and $\kappa$ along with two velocity parameters $\alpha_{\mathrm{cen}}$ and $\alpha_{\mathrm{sat}}$, making a total of seven parameters:
\begin{align}
    \{ M_{\mathrm {cut}},M_1,\sigma,\alpha,\kappa, \alpha_{\rm{cen}},\alpha_{\rm{sat}} \}\,.
\end{align}
There have been several studies exploring the impact of assembly bias, which considers how secondary halo properties beyond halo mass can influence galaxy occupation \cite{Wechsler2002ConcentrationsHistories,Gao2005TheClustering, Croton2007HaloClustering, Gao2007AssemblyHaloes, Lin2016OnPopulations,Pujol2017WhatDensity, Artale2018TheSimulations, Zehavi2018TheHalos, Hadzhiyska2020LimitationsBeyond, Xu2021DissectingBias, Xu2021PredictingLearning, Delgado2022ModellingLearning, Salcedo2022ElucidatingSDSS, Yuan2022IllustratingILLUSTRISTNG, Wang2022EvidenceStatistics, Yuan2023UnravelingData}. However, in this analysis, we choose to use the baseline HOD prescription described above and leave extensions to galaxy halo connection model for future work. Although, given the scales we explored, we do not expect the results to show significant differences, since assembly bias have more obvious effect on smaller scales.

With the HOD prescription described above, we can generate real-space mock galaxy catalogues based on dark matter only simulations. To transform these into redshift space, we apply RSD to the mock galaxy positions as follow:
\begin{align}
    \mathbf{r}_{\mathrm{redshift}} = \mathbf{r}_{\mathrm{real}} + \frac{\left( \mathbf{v}_{\mathrm{pec}} \cdot \hat{\mathbf{r}}_{\mathrm{LoS}} \right)(1+z)}{H(z)} \hat{\mathbf{r}}_{\mathrm{LoS}}\,,
    \label{eqn:rsd}
\end{align} 
where $ \mathbf{r}_{\mathrm{real}}$ is the real-space position vector of the galaxy, $ \mathbf{v}_{\mathrm{pec}}$ is the peculiar velocity vector $\hat{\mathbf{r}}_{\mathrm{LoS}}$ is the unit vector in the direction of the LoS, and $H(z)$ is the Hubble parameter, with the $1+z$ factor converting to comoving units.

\subsection{Galaxy power spectrum in the EFTofLSS}
\label{subsec:eft}

The EFTofLSS provides a robust framework for modeling the redshift-space galaxy power spectrum by systematically incorporating the effects of small-scale physics on large-scale clustering. EFTofLSS extends traditional perturbation theory by including additional counterterms that account for the influence of small-scale physics, including the galaxy formation processes. We briefly summarize the model here and refer readers to \cite{Perko2016BiasedStructure,dAmico2020TheStructure} for further details.

The model for the redshift space galaxy power spectrum at one loop EFT is given by:

\begin{align}\label{eqn:gpk}
 P_{g}(k, \mu) & = Z_1(\mu)^2 P_{11}(k) + 2 \int \frac{d^3q}{(2\pi)^3}\; Z_2(\mathbf{q},\mathbf{k}-\mathbf{q},\mu)^2 P_{11}(|\mathbf{k}-\mathbf{q}|)P_{11}(q)\nonumber  \\
& + 6 Z_1(\mu) P_{11}(k) \int\, \frac{d^3 q}{(2\pi)^3}\; Z_3(\mathbf{q},-\mathbf{q},\mathbf{k},\mu) P_{11}(q)\nonumber \\
& + 2 Z_1(\mu) P_{11}(k)\left( c_\text{ct}\frac{k^2}{{ k^2_\textsc{m}}} + c_{r,1}\mu^2 \frac{k^2}{k^2_\textsc{r}} + c_{r,2}\mu^4 \frac{k^2}{k^2_\textsc{r}} \right)\nonumber \\
& + \frac{1}{\bar{n}_g}\left( c_{\epsilon,0}+c_{\epsilon,1}\frac{k^2}{k_\textsc{m}^2} + c_{\epsilon,2} f\mu^2 \frac{k^2}{k_\textsc{m}^2} \right) \,,
\end{align}
Our notation follows that established in \cite{DAmico2021LimitsCode}, incorporating a combination of linear terms, 1-loop Standard Perturbation Theory (SPT) terms, counterterms, and stochastic terms. The quantity $\mu$ is the cosine of the angle between the LoS direction and wavenumber vector $\mathbf{k}$. $P_{11}(k)$ denotes the linear matter power spectrum, and $f$ is the growth factor. The scale $k_\textsc{m}^{-1}$ controls the spatial extension of the collapsed objects, the scale $k_\textsc{r}^{-1}$ controls the counterterms needed to define the product of velocities at the same point~\cite{DAmico2024TamingData,Ivanov2022CosmologicalDistortions}, and $\bar{n}_g$ is the mean galaxy number density.
In the following, we will set $k_\textsc{m}^{-1} = 0.7  \, h^{-1} \rm {Mpc}$ and $k_\textsc{r}^{-1} = 0.35  \, h^{-1} \rm {Mpc}$, while $\bar{n}_g$ is estimated for each box.
$Z_n$ represents the $n$th order of the redshift galaxy density kernel,
\begin{align}\label{eqn:zkernels}
    Z_1(\mathbf{q}_1) & = K_1(\mathbf{q}_1) +f\mu_1^2 G_1(\mathbf{q}_1) = b_1 + f\mu_1^2\,, \nonumber\\ 
    Z_2(\mathbf{q}_1,\mathbf{q}_2,\mu) & = K_2(\mathbf{q}_1,\mathbf{q}_2) +f\mu_{12}^2 G_2(\mathbf{q}_1,\mathbf{q}_2)+ \, \frac{1}{2}f \mu q \left( \frac{\mu_2}{q_2}G_1(\mathbf{q}_2) Z_1(\mathbf{q}_1) + \text{perm.} \right)\,, \nonumber\\ 
    Z_3(\mathbf{q}_1,\mathbf{q}_2,\mathbf{q}_3,\mu) & = K_3(\mathbf{q}_1,\mathbf{q}_2,\mathbf{q}_3) + f\mu_{123}^2 G_3(\mathbf{q}_1,\mathbf{q}_2,\mathbf{q}_3) \nonumber \\ 
    &+ \frac{1}{3}f\mu q \left(\frac{\mu_3}{q_3} G_1(\mathbf{q}_3) Z_2(\mathbf{q}_1,\mathbf{q}_2,\mu_{123}) +\frac{\mu_{23}}{q_{23}}G_2(\mathbf{q}_2,\mathbf{q}_3)Z_1(\mathbf{q}_1)+ \text{cyc.}\right)\,,
\end{align}
where $K_n$ are $n$-th order galaxy density kernels,
\begin{align}\label{eqn:kkernels}
    K_1 & = b_1\,, \nonumber\\
    K_2(\mathbf{q}_1,\mathbf{q}_2) & = b_1\frac{\mathbf{q}_1\cdot \mathbf{q}_2}{q_1^2}+ b_2\left( F_2(\mathbf{q}_1,\mathbf{q}_2)- \frac{\mathbf{q}_1\cdot \mathbf{q}_2}{q_1^2} \right) + b_4 + \text{perm.} \,, \nonumber\\
    K_3(k, q) & = \frac{b_1}{504 k^3 q^3}\left( -38 k^5q + 48 k^3 q^3 - 18 kq^5 + 9 (k^2-q^2)^3\log \left[\frac{k-q}{k+q}\right] \right) \nonumber\\
    &+ \frac{b_3}{756 k^3 q^5} \left( 2kq(k^2+q^2)(3k^4-14k^2q^2+3q^4)+3(k^2-q^2)^4 \log \left[\frac{k-q}{k+q}\right] \nonumber \right) \,.
\end{align}
The explicit expressions for the second-order density kernel $F_2$ and velocity kernels $G_n$ of the standard perturbation theory are omitted here for brevity. To summarize, we use four bias parameters: $b_1, b_2, b_3$, and $b_4$; three parameters for counterterms: $c_{\rm{ct}}, c_{r,1}$, and $c_{r,2}$; and three parameters for stochastic terms $c_{\epsilon,0}, c_{\epsilon,1}, c_{\epsilon,2}$. This gives a total of 10 parameters:
\begin{align}
    \{ b_1, b_2, b_3, b_4, c_{\rm{ct}}, c_{r,1}, c_{r,2}, c_{\epsilon,0}, c_{\epsilon,1}, c_{\epsilon,2} \}\,.
\end{align}
The theoretical power spectrum is properly IR-resummed \cite{Senatore2014RedshiftStructures, Senatore2015TheStructures, Lewandowski2020AnPeak} to handle long-wavelength modes that could otherwise lead to divergences in perturbative calculations. 

\subsection{Alcock-Paczynski distortions}
\label{subsec:ap}
When we fit to data, we apply the Alcock-Paczynski(AP) transformation to account for potential distortions arising when converting redshifts to distances, due to the offset between the fiducial cosmology and the true cosmology. We use the standard method for addressing these distortions, where the perpendicular and parallel distortion factors can be computed as:
\begin{equation}
    \begin{aligned}
    q_{\parallel} &= \frac{H^{\rm fid}(z)/H_0^{\rm fid}}{H(z)/H_0}\,,\\
    q_{\perp} &= \frac{D_\mathrm{A}(z)H_0}{D^{\rm fid}_\mathrm{A}(z)H_0^{\rm fid}}\,,
    \end{aligned}
\end{equation}
where $H(z)$ and $D_\mathrm{A}(z)$ are the Hubble parameter and the angular diameter distance at redshift $z$, respectively. The superscript \texttt{fid} denotes the quantities computed in the fiducial cosmology used for converting redshifts to distances, while non-superscripted quantities refer to those computed in the observed cosmology. The magnitude and angle of the wavenumber computed in the model (primed quantities) are then related to the observed magnitude and angle (unprimed quantities) by:
\begin{equation}\label{eq:scaling}
\begin{aligned}
k' &= \frac{k}{q_{\perp}}\left[1 + \mu^2\left(\frac{1}{F^2} - 1\right)\right]^{1/2}\,,
\\
\mu' &= \frac{\mu}{F_{}}\left[1 + \mu^2\left(\frac{1}{F^2} - 1\right)\right]^{-1/2}\,,
\end{aligned}
\end{equation}
where $F = q_{\parallel}/q_{\perp}$. The modeled multipole power spectrum with AP effect then read: 
\begin{equation}\label{eq:multi}
\begin{aligned}
P_{\rm \ell}(k) &= \frac{(2\ell + 1)}{2q^2_{\perp}q_{\parallel}}\int^1_{-1}d\mu\; P_{g}\left[k'(k, \mu), \mu'(\mu)\right]\mathcal{L}_{\ell}(\mu)\,,
\end{aligned}
\end{equation}
where the power spectrum under the integral can be computed using eq.~\ref{eqn:gpk}, and $\mathcal{L}_{\ell}(\mu)$ is the Legendre polynomials. 

\section{Methodology}
\label{sec:method}

The primary motivation for this paper is to establish a connection between the HOD model and the EFT parameter space, with the goal of effectively reducing the nuisance parameter space volume of the EFT model, using physical constraints. By linking these two frameworks, we can leverage the strengths of the HOD model to inform the EFT parameters, resulting in a more precise and efficient cosmological analysis.

To achieve this, we outline our comprehensive pipeline in the following subsections. First, we generate the galaxy power spectrum using a broad range of HOD parameters from N-body simulations across different cosmologies, which we detail in Section~\ref{subsec:poppk}. Section~\ref{subsec:fiteft} then explains the method for rapidly estimating the best-fit EFT parameters for each power spectrum measurement. The distribution of these best-fit EFT parameters forms our HOD-informed EFT parameter distribution. In Section~\ref{subsec:nf}, we describe how a normalizing flow is employed to better understand this distribution and to generate HOD-informed prior likelihoods for arbitrary sets of EFT parameters, which can be subsequently used in cosmological analyses.

\subsection{Generating galaxy power spectra from HOD and N-body simulations}
\label{subsec:poppk}

We use the \textsc{AbacusSummit} simulation suite \cite{Garrison2016ImprovingSimulations,Garrison2018TheSimulations,Garrison2019AABACUS,Garrison2021TheCode,Bose2022ConstructingABACUSSUMMIT} for this analysis, which provides a set of large, high-resolution N-body simulations specifically designed to meet the accuracy requirements of the DESI galaxy survey. These simulations cover a wide range of cosmological parameters, making them particularly suitable for our study. For this work, we utilize the \texttt{AbacusSummit\char`_base\char`_c000\char`_ph000} box, which represents the Planck 2018 cosmology \cite{Collaboration2020PlanckParameters}, along with the \texttt{AbacusSummit\char`_base\char`_c130\char`_ph000} to \texttt{AbacusSummit\char`_base\char`_c160\char`_ph000} boxes at $z=0.8$. The latter form an unstructured emulator grid surrounding the \texttt{c000} box, providing wider coverage of the 7-dimensional cosmological space $\{ \omega_b h^2, \omega_c h^2, h, A_s, n_s, w_0, w_a \}$. These 32 boxes were specifically selected to collectively sample variations in the cosmological parameters of interest, ensuring broad and scattered coverage across the parameter space. Each simulation box has a size of $2000 \, h^{-1} \rm {Mpc}$ per side and contains $6912^3$ dark matter particles. This setup allows us to generate galaxy power spectra with diverse cosmological parameters, ensuring generality in our findings - the prior should not be tied to a single comsological model, which might then bias results. The cosmological dependencies of the resulting HOD-informed EFT parameter distribution will be discussed in detail in Section~\ref{subsec:cosmo_dp}.

\begin{table}[t]
\centering
\label{table:hod_parameters}
\begin{tabular}{|c|c|}
\hline
\textbf{Parameter} & \textbf{Distribution} \\ \hline
$\log_{10} M_{\mathrm{cut}}$ &  $\mathcal{N}(\mu=13.0, \sigma=1)$, truncated $[12.0, 13.8]$ \\
$\log_{10} M_1$ & $\mathcal{N}(\mu=14.0, \sigma=1)$, truncated $[12.5, 15.5]$ \\ 
$\sigma$ & $\mathcal{N}(\mu=0.5, \sigma=0.5)$, truncated $[0.0, 3.0]$ \\ 
$\alpha$ & $\mathcal{N}(\mu=1.0, \sigma=0.5)$, truncated $[0.0, 2.0]$ \\ 
$\kappa$ & $\mathcal{N}(\mu=0.5, \sigma=0.5)$, truncated $[0.0, 3.0]$ \\ 
$\alpha_{\rm{cen}}$ & $\mathcal{N}(\mu=0.4, \sigma=0.4)$, truncated $[0.0, 1.0]$ \\
$\alpha_{\rm{sat}}$ & $\mathcal{N}(\mu=0.8, \sigma=0.4)$, truncated $[0.0, 2.0]$ \\ \hline
\end{tabular}
\caption{HOD sample distributions for the seven HOD parameters. Each parameter is sampled from a normal distribution with truncation applied where specified. We also test the impact of removing the normal distribution, leaving only the truncation; see Section~\ref{sec:prior} for details.}
\label{tab:hoddist}
\end{table}

The dark matter halos are identified using the \textsc{CompaSO} halo finder, a highly efficient, on-the-fly group finder specifically designed for the \textsc{AbacusSummit} simulations \cite{Hadzhiyska2022COMPASO:Overdensities}. We use the \textsc{AbacusHOD} code \cite{Yuan2022ABACUSHOD:Data} to rapidly generate 10,000 power spectrum multipole measurements for each simulation box, corresponding to 10,000 different sets of HOD parameters. These measurements are produced using the HOD model described in Section~\ref{subsec:hod}, with the HOD parameters sampled according to the distributions listed in Table~\ref{tab:hoddist}. The HOD parameter distributions are broad and based on the priors selected for the DESI LRG HOD study by \cite{Yuan2024TheABACUSSUMMIT}, providing a conservative choice to inform the EFT parameter space. Unlike HOD fitting, our approach does not aim to imprint small-scale clustering information into the HOD-informed EFT parameter space. Instead, by using a broad range of HOD parameters and varying cosmologies, we ensure that the resulting HOD-informed EFT parameter distribution is primarily shaped by the structure of the HOD model and the selected HOD distributions for specific types of galaxies. It is important to note that the choice of HOD model and parameter distributions could impact the resulting EFT parameter space, as it directly reflects the characteristics of the galaxies being studied. However, we find that the results are insensitive to the choice of parameter distribution, suggesting the primary impact of the HOD prior is to remove unphysical regions of EFT parameter space. A detailed discussion of the HOD parameter distribution choices will be presented in Section~\ref{subsec:hod_dp}, highlighting how they shape the EFT parameter landscape. In future work, we will apply this method to the DESI data using multiple HOD models and parameter choices, further testing this dependence.

\subsection{Rapid best-fit estimation of EFT parameters}
\label{subsec:fiteft}
 
We perform EFT parameter fits to the galaxy power spectrum multipoles generated from 32 simulation boxes, using the theoretical model described in Section~\ref{subsec:eft}. We use \textsc{Effort.jl}\footnote{\url{https://github.com/CosmologicalEmulators/Effort.jl}} \cite{Bonici:2025ltp} to compute the galaxy clustering multipoles, as a function of cosmological and nuisance parameters. \textsc{Effort.jl}  is a novel emulator for the EFTofLSS based on \textsc{PyBird}\footnote{\url{https://github.com/pierrexyz/pybird}} \cite{DAmico2021LimitsCode}, which shares the same computational backend of \textsc{Capse.jl}~\cite{Bonici:2023xjk} and whose main feature is its compatibility with the \textsc{Turing.jl}\footnote{\url{https://github.com/TuringLang/Turing.jl}} \cite{Ge2018Turing:Inference} library and automatic differentiation engines. \textsc{Turing.jl} is a probabilistic programming language which can be used to define differentiable likelihoods and is interfaced with gradient-based minimizers such as the Limited memory Broyden–Fletcher–Goldfarb–Shanno (L-BFGS) algorithm \cite{Liu1989OnOptimization}, which ensures a quick and reliable convergence to the maximum of the log-likelihood.

We fit the monopole and quadrupole up to $k_{\rm{max}} = 0.2\, {\rm{Mpc}}^{-1} h$ as our baseline results, with the impact of including the hexadecapole discussed in Section~\ref{subsec:hod_dp}.
The choice of $k_{\rm{max}} = 0.2\, {\rm{Mpc}}^{-1} h$ is reasonable given the samples and the volumes considered in the tests in Section~\ref{sec:comparison}; we thus match the same $k_{\rm{max}}$ in the best-fitting procedure.
When the hexadecapole is not included, the parameters $c_\mathrm{r,2}$ and $c_{\epsilon,1}$ are fixed to 0, as $c_\mathrm{r,2}$ is exactly degenerate with $c_\mathrm{r,1}$ and $c_{\epsilon,1}$ could not be well constrained by just the monopole and quadrupole. When the hexadecapole is included, we continue to use $k_{\rm{max}} = 0.2\, {\rm{Mpc}}^{-1} h$ and allow all ten EFT parameters to vary. During the best-fit finding process, we fix the cosmology to match the corresponding simulation box and vary only the EFT parameters. Additionally, we set the model galaxy density to match the true galaxy density corresponding to the power spectrum we fit to, in order to consistently fit the stochastic coefficients across all samples.
Given the broad range of the HOD parameter space, we use wide, uninformative flat priors for the best-fit estimation at this stage of the process. The broad priors used, along with the standard priors for EFTofLSS-based full-shape cosmological analysis, are listed in Table~\ref{tab:eftprior}.

Finally, we run the likelihood maximization procedure eight times for each HOD model, starting from different initial guesses and retaining the best solution. This approach helps avoiding convergence to local minima and ensures the robustness of our results.

\begin{table}[t]
\centering
\begin{tabular}{|c|c|c|}
\hline
\textbf{Parameter} & \textbf{Best fit finding Prior} & \textbf{Standard prior} \\ \hline
$b_1$ & $\mathcal{U}(0.0, 12.0)$ & $\mathcal{U}(0.0, 4.0)$ \\
$b_2$ & $\mathcal{U}(-24.0, 24.0)$ & $\mathcal{U}(-4.0, 4.0)$ \\
$b_3$ & $\mathcal{U}(-90.0, 90.0)$ & $\mathcal{N}(0.0, 10.0)$ \\
$b_4$ & $\mathcal{U}(-24.0, 24.0)$ & $\mathcal{N}(0.0, 2.0)$ \\
$c_{\rm{ct}}$ & $\mathcal{U}(-36.0, 36.0)$ & $\mathcal{N}(0.0, 4.0)$ \\
$c_{r,1}$ & $\mathcal{U}(-72.0, 72.0)$ & $\mathcal{N}(0.0, 8.0)$ \\
$c_{r,2}$ & $\mathcal{U}(-72.0, 72.0)$ & $\mathcal{N}(0.0, 8.0)$ \\
$c_{\epsilon,0}$ & $\mathcal{U}(-24.0, 24.0)$ & $\mathcal{N}(0.0, 2.0)$ \\
$c_{\epsilon,1}$ & $\mathcal{U}(-90.0, 90.0)$ & $\mathcal{N}(0.0, 4.0)$ \\
$c_{\epsilon,2}$ & $\mathcal{U}(-72.0, 72.0)$ & $\mathcal{N}(0.0, 4.0)$ \\ \hline
\end{tabular}
\caption{EFT parameter priors for best fit finding and standard EFT-based full-shape analysis}
\label{tab:eftprior}
\end{table}

For our analysis, we use the \textsc{CovaPT} code\footnote{\url{https://github.com/JayWadekar/CovaPT}} to generate a Gaussian analytic covariance matrix based on perturbation theory \cite{Wadekar2020GalaxyTheory}. It is important to note that the resulting HOD-informed EFT parameter distribution is not strongly dependent of the specific covariance used. Further details on this independence are provided in Appendix~\ref{app:cov_dp}.

\subsection{Learning the distribution with normalizing flow}
\label{subsec:nf}

To generate a HOD-informed prior that covers arbitrary inputs of EFT parameter sets within the exploring range, we employ a Normalizing Flow (NF) model \cite{JimenezRezende2015VariationalFlows}. We begin by compiling the best-fit EFT parameters obtained from the 32 simulation boxes. Outliers that are close to the boundaries defined in Table~\ref{tab:eftprior} are removed to ensure robust training data, accounting for approximately 5$\%$ of the total samples. We pre-process the data using a min-max scaling strategy to normalize the input features. The scaled data is then split into training and validation sets, with 70$\%$ of the data used for training and the remaining 30$\%$ for validation. This split ensures that the model generalizes well to unseen data.

The flows are implemented using the \textsc{nflows}\footnote{\url{https://github.com/bayesiains/nflows}} \cite{Durkan2019NeuralFlows} library, which provides the flow transformations for the NF model. The model structure and training process are managed using \textsc{PyTorch Lightning}\footnote{see \url{https://github.com/pytorch/pytorch} for \textsc{PyTorch} and \url{https://github.com/Lightning-AI/pytorch-lightning} for \textsc{PyTorch Lightning}} \cite{Paszke2019PyTorch:Library}, which simplifies the setup and ensures scalability. The NF model is built with 10 features, 10 layers of Masked Affine Autoregressive Transforms, and a batch size of 256 for efficient training. The base distribution is a standard normal distribution, and the transforms include reverse permutations to enhance the model's expressiveness. We use the Adam optimizer \cite{Kingma2014Adam:Optimization} with a learning rate of 0.005 during training. The training process minimizes the negative log-likelihood of the data under the flow model, which corresponds to maximizing the probability of the observed data. We log the training and validation losses to monitor performance, and the best models are saved based on the validation loss using a checkpoint callback. The NF model effectively learns the distribution of the best-fit EFT parameters, providing a reliable HOD-informed prior. In the next section, we will demonstrate the robustness of this approach and present the resulting learned prior.

\section{HOD-informed priors}
\label{sec:prior}

We now analyze the behavior of the HOD-informed prior under various conditions. We begin by testing its sensitivity to different cosmological models and demonstrating the effectiveness of the NF model in learning the distribution. We then investigate the impact of both the HOD sample distributions and the inclusion of the hexadecapole moment on the resulting EFT parameter space. 
We set three configurations to evaluate their effects on the prior formation:
\begin{itemize}
    \item TNHOD M+Q: This configuration uses truncated normal HOD samples as described in Table~\ref{tab:hoddist}, incorporating only the monopole and quadrupole moments. It serves as the baseline configuration for our analysis, with 8 free EFT parameters ($c_\mathrm{r,2}$ and $c_{\epsilon,1}$ fixed to 0).
    \item UHOD M+Q: In this configuration, the HOD samples are uniformly distributed but share the same truncation limits as the TNHOD configuration, omitting the Gaussian component. Like the baseline, only the monopole and quadrupole are used, with 8 free EFT parameters.
    \item TNHOD M+Q+H: This configuration extends the baseline by including the hexadecapole moment, providing additional information for prior formation. In this case, all 10 EFT parameters are free to vary.
\end{itemize}

We also perform a perturbative check to ensure that the HOD-informed EFT parameter space satisfies the perturbative requirements. Further details are provided in the Appendix~\ref{app:pert_check}.

\subsection{Cosmology dependency and robustness of NF}
\label{subsec:cosmo_dp}

\begin{figure}
    \centering
    \includegraphics[width=0.80\linewidth]{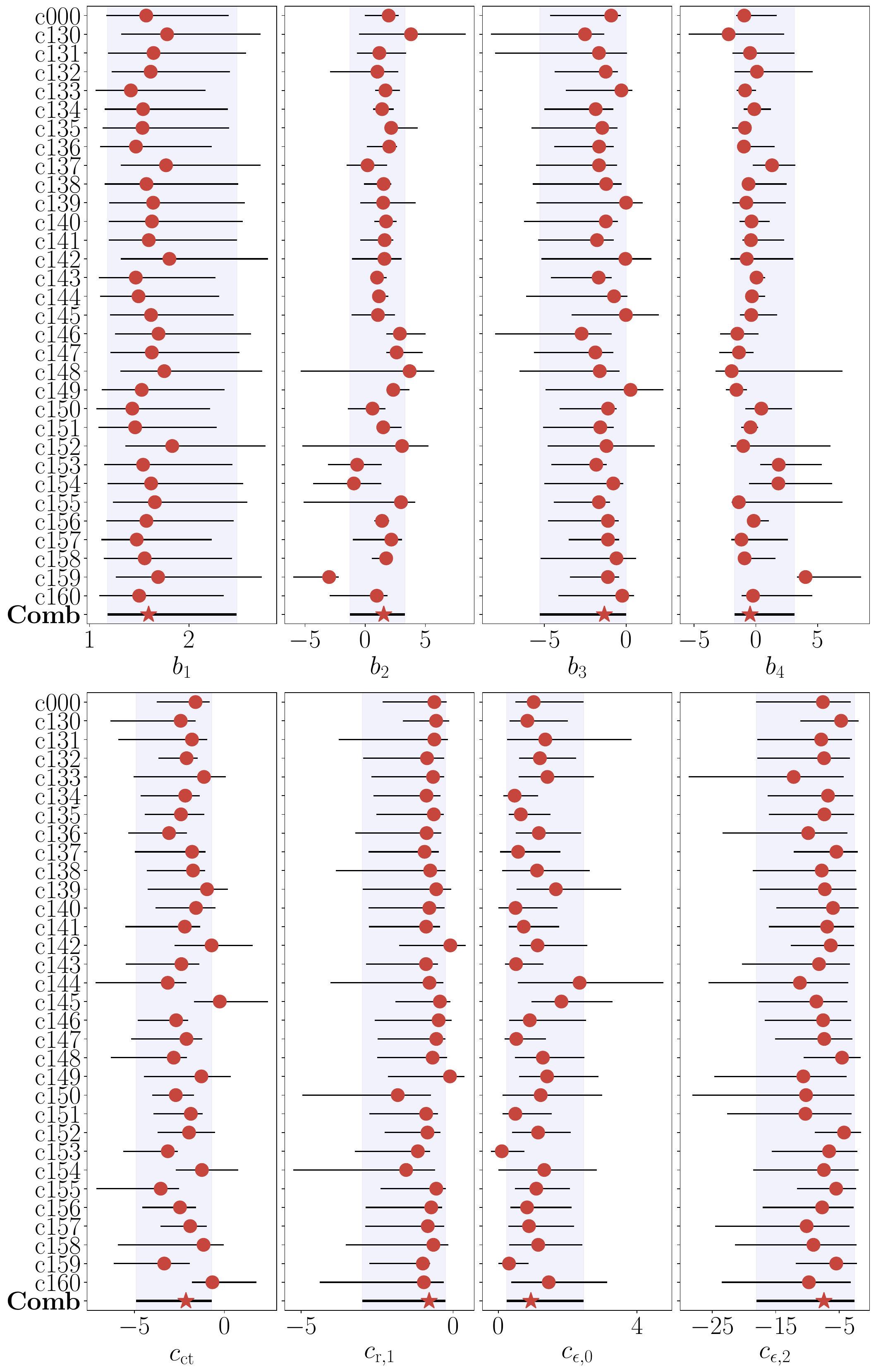}
    \caption{The median and 68th percentile of the best-fit EFT parameter distributions for the TNHOD M+Q configuration, prior to applying the NF model. Each panel represents one of the eight EFT parameters across 32 individual simulation boxes, with the combined case (Comb) shown at the bottom. The red points represent the median values for each box, and the horizontal error bars indicate the 68th percentile range of the best-fit distributions. The shaded regions represent the 68th percentile for the combined case.}
    \label{fig:cosmo_dp}
\end{figure}

\begin{figure}
    \centering
    \includegraphics[width=0.95\linewidth]{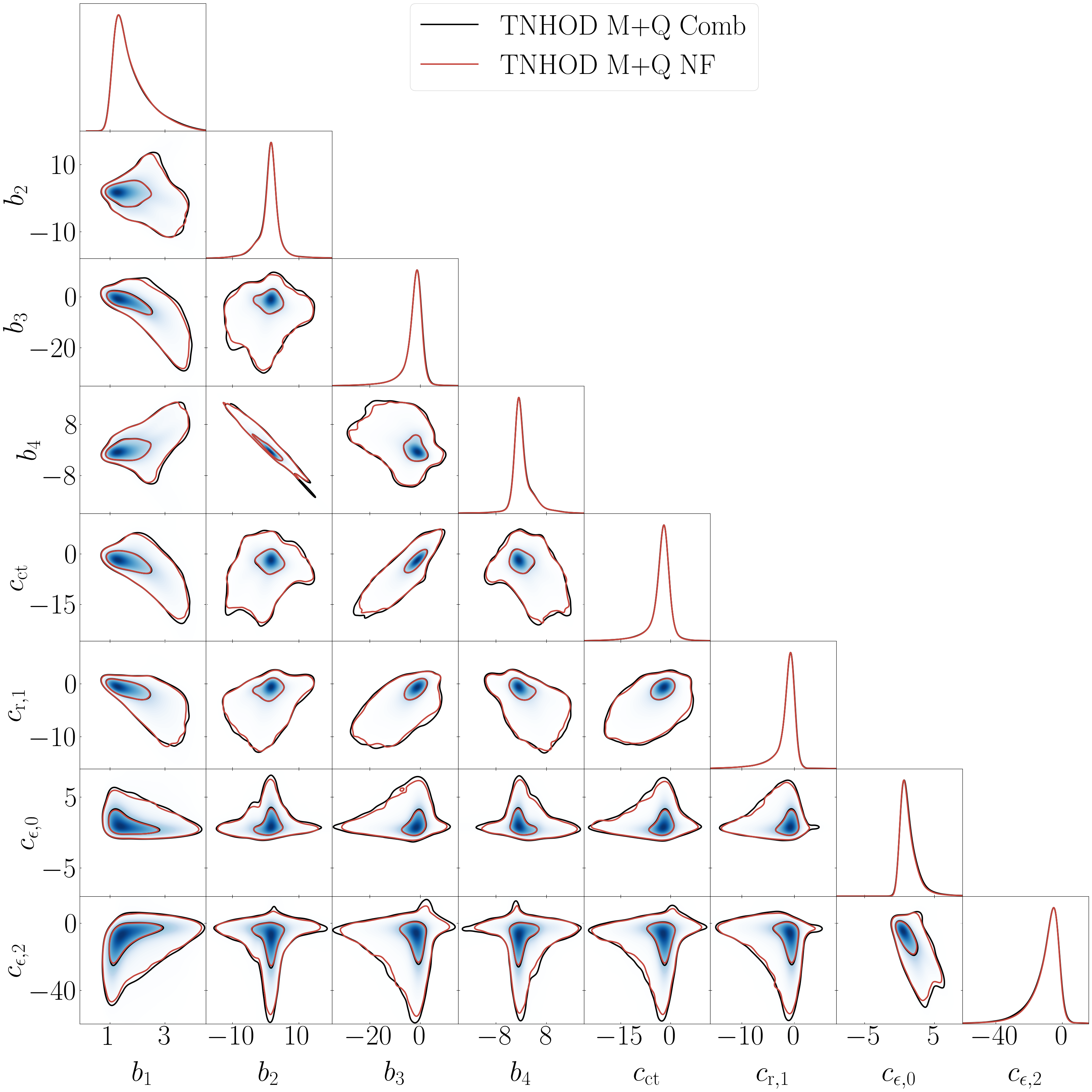}
    \caption{A comparison between the original combined best-fit EFT parameter distributions (black shaded contours) and the distributions learned by the NF model (red contours) for the TNHOD M+Q configuration. The diagonal panels show the 1D marginalized distributions, while the off-diagonal panels display the 1-$\sigma$ and 2-$\sigma$ contours.}
    \label{fig:tri_comb_nf}
\end{figure}

Figure~\ref{fig:cosmo_dp} displays the median and 68th percentile of the best-fit distribution under the TNHOD M+Q configuration, prior to applying the NF, for each individual simulation box as well as the combined case (Comb) across all 32 boxes. Each red point represents the median value of an EFT parameter, with the error bars denoting the 68th percentile range. The consistency between the individual boxes and the combined result is apparent for all EFT parameters explored, with most individual box distributions overlapping well with the combined case. This indicates that no strong cosmological dependency is observed in the EFT parameter distributions, ensuring the safety of using the HOD-informed prior when exploring the cosmological parameter space.

Figure~\ref{fig:tri_comb_nf} compares the true combined best-fit EFT parameter distributions (shaded black contours) to those learned by the NF model (red contours) for the TNHOD M+Q configuration. The 1D marginalized distributions and 1-$\sigma$ contours show excellent alignment between the true and NF-learned distributions. Only slight differences appear in the 2-$\sigma$ contour lines, demonstrating the NF model's robustness in capturing the distribution, with minimal discrepancies at the outer edges. 

\subsection{Influence of HOD sample distribution and hexadecapole inclusion}
\label{subsec:hod_dp}

\begin{figure}
    \centering
    \includegraphics[width=0.95\linewidth]{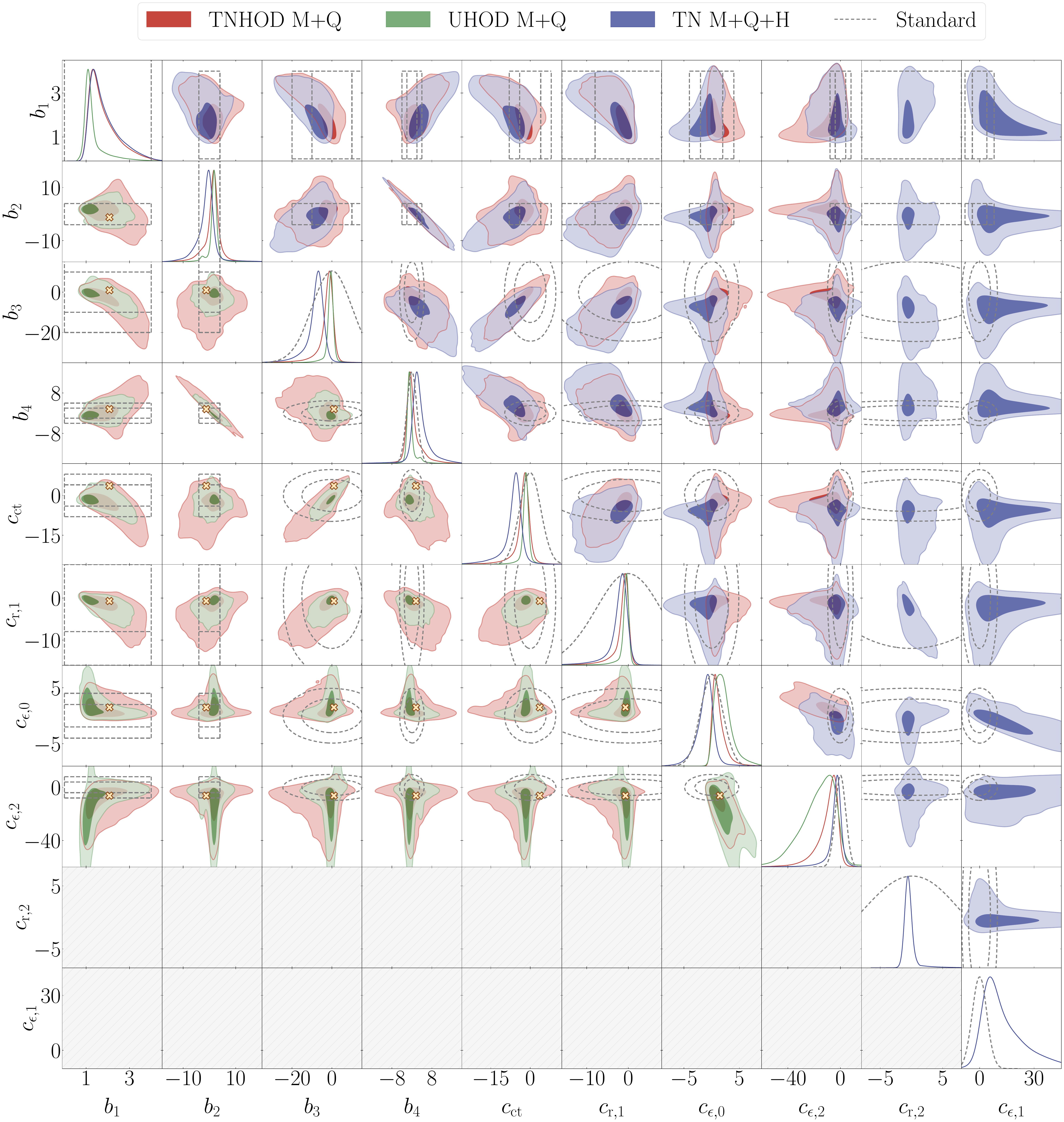}
    \caption{Contour plots illustrating the NF learned HOD-informed priors using three different configurations: TNHOD M+Q (red contours), UHOD M+Q (green contours), and TNHOD M+Q+H (blue contours). The unfilled gray contours (and boundary lines) represent the standard prior for EFTofLSS full-shape analysis as listed in Table~\ref{tab:eftprior} for comparison. Each contour indicates the 1-$\sigma$ and 2-$\sigma$ intervals of the distribution. The lower triangle compares the TNHOD M+Q and UHOD M+Q priors, highlighting the impact of removing the Gaussian component from the HOD sample distribution. Cross markers indicate the best-fit EFT parameters for the DESI-like Abacus mock power spectrum (monopole and quadrupole) used in Section~\ref{sec:comparison}, with cosmological parameters fixed to their true values. The upper triangle compares the TNHOD M+Q and TNHOD M+Q+H priors, showing the effect of including the hexadecapole moment on the priors.}
    \label{fig:hexa_hod_dp}
\end{figure}

The NF-learned HOD-informed priors for the three configurations are displayed in Figure~\ref{fig:hexa_hod_dp}. In this figure, the red contours correspond to the TNHOD M+Q configuration, the green contours to the TNHOD M+Q+H configuration, and the slate blue contours to the UHOD M+Q configuration. For comparison, the standard EFTofLSS full-shape analysis prior, as listed in Table~\ref{tab:eftprior}, is shown as a gray dashed contour.

Focusing on the lower triangle in this figure, we compare the impact of removing the Gaussian prior when sampling the HOD parameters. The most notable difference is observed in $b_1$, where the UHOD configuration shows a more concentrated $b_1$ distribution, indicated by a smaller 1-sigma contour area and 1D marginalized distribution. This occurs because $b_1$ is the most degenerate parameter with the HOD parameters. Specifically, in the UHOD configuration, we sample more combinations with large $\sigma$ and smaller $\log M_\mathrm{cut}$. These combinations tend to lead to smaller values of $b_1$.

From the HOD model perspective, larger $\sigma$ means that the transition from low to high probability of a halo hosting a galaxy is less steep, allowing a broader range of halo masses to contribute to the galaxy population. This more gradual transition results in a higher number of low-mass halos hosting galaxies, which generally have lower bias, thus reducing the overall galaxy bias. Similarly, a smaller $\log M_\mathrm{cut}$ lowers the threshold mass for halos to begin hosting galaxies, further increasing the contribution from low-mass halos. These low-mass halos, with lower bias, also contribute to reducing the overall value of $b_1$. Since $b_1$ represents the linear galaxy bias, its value is closely tied to the mass distribution and occupancy of galaxies within halos. The inclusion of more extreme HOD parameter combinations under the UHOD configuration therefore skews the $b_1$ distribution towards smaller values, reflecting these physical relationships.

Despite the differences in the distribution of $b_1$, the overall volume of the parameter space is smaller for the UHOD M+Q configuration compared to TNHOD M+Q, as indicated by the generally smaller 2-$\sigma$ contour sizes in most cases. As discussed earlier for $b_1$, the UHOD configuration allows for more galaxies to reside in low-mass host halos, leading to a different galaxy population compared to TNHOD. Consequently, the two configurations represent different types of galaxies, with TNHOD being more LRG-like, reflecting galaxies in more massive halos, while UHOD produces a more concentrated best-fit EFT distribution. However, aside from $b_1$, the peak values of the 1D distributions remain largely unchanged for most other parameters.

Next, we focus on the upper triangle in Figure~\ref{fig:hexa_hod_dp}, where we compare the impact of including the hexadecapole. 
For the M+Q cases, we set $c_\mathrm{r,2}$ to 0, as it is degenerate with $c_\mathrm{r,1}$. However, when we include the hexadecapole, $c_\mathrm{r,2}$ can be well constrained. 
Additionally, examining the 1D marginalized distributions, we observe slight shifts in several parameters with the inclusion of the hexadecapole. 
The higher-order bias terms $b_2$, $b_3$ and $b_4$ are affected as the hexadecapole provide more information about the anisotropic clustering of galaxies. 
The counterterm coefficient $c_\mathrm{ct}$ which account for corrections to non-linear effects, is influenced by the hexadecapole’s ability to distinguish velocity dispersion effects from galaxy clustering anisotropies.
$c_{\epsilon,0}$ tends to favor smaller values due to the inclusion of $c_{\epsilon,1}$ as free parameter.
Nevertheless, the distribution remains largely consistent between the M+Q and M+Q+H cases.

When comparing all HOD-informed priors with the standard prior, it is clear that the HOD-informed priors, regardless of the configuration, provide a detailed picture of the degeneracy directions between the EFT parameters, which the standard prior fails to capture. Additionally, for most parameters, the 1-$\sigma$ contour areas are significantly smaller for the HOD-informed priors, indicating a noticeable reduction in the effective EFT parameter space. This reduction shows that the HOD-informed priors help guide the exploration of a more physically meaningful parameter space, mitigating prior effects and ultimately leading to more accurate cosmological parameter posteriors. Since our power spectrum measurements are derived from finite-volume simulation boxes, the best-fit EFT parameters obtained from any single measurement will scatter around the ``true'' values that would be expected in an infinite-volume limit. This additional scatter, primarily driven by sample variance, effectively acts as a Gaussian smoothing on the HOD-informed priors, broadening them and making them more conservative. The priors still account for HOD variations and robustly capture the most plausible EFT parameter values, while conservatively excluding unphysical regions.

\section{Comparative analysis of priors}  \label{sec:comparison}

\begin{table}
\resizebox{\columnwidth}{!}{%
\begin{tabular}{|p{40pt}|*{3}{>{\centering\arraybackslash\hspace{0pt}}p{48pt}|>{\centering\arraybackslash\hspace{0pt}}p{48pt}|>{\centering\arraybackslash\hspace{0pt}}p{35pt}|}}
\cline{1-10}
Model & \multicolumn{3}{c|}{$\Lambda$CDM} &
\multicolumn{3}{c|}{$w$CDM} &
\multicolumn{3}{c|}{$w_0 w_a$CDM}\\
\cline{1-10}
Prior &Standard&TNHOD&UHOD&Standard&TNHOD&UHOD&Standard&TNHOD&UHOD\tabularnewline
\cline{1-10}
$\Omega_c h^2$      &1.65(2.46) &0.57(1.36) &0.46 &1.90(2.64) &1.03(1.44) &0.96 &2.00(2.60) &1.00(1.39) &1.04
\tabularnewline
$H_0$               &1.55(1.74) &0.87(1.08) &0.82 &2.26(1.36) &1.57(0.88) &1.56 &1.84(0.93) &1.76(1.04) &1.38
\tabularnewline
$\ln 10^{10} A_s$   &1.46(1.61) &0.37(0.74) &0.22 &2.55(1.95) &1.55(0.81) &1.42 &2.89(1.61) &2.05(0.90) &1.72
\tabularnewline
$n_s$               &1.88(2.12) &0.50(1.29) &0.32 &1.97(2.31) &1.13(1.35) &0.96 &2.09(2.31) &1.17(1.41) &1.06
\tabularnewline
$w$                 &\hspace{7.91pt}-\hspace{7.91pt}(\hspace{7.91pt}-\hspace{7.91pt}) &\hspace{7.91pt}-\hspace{7.91pt}(\hspace{7.91pt}-\hspace{7.91pt}) &- &1.97(0.80) &1.42(0.44) &1.40 &0.72(0.21) &1.01(0.56) &0.64
\tabularnewline
$w_a$               &\hspace{7.91pt}-\hspace{7.91pt}(\hspace{7.91pt}-\hspace{7.91pt}) &\hspace{7.91pt}-\hspace{7.91pt}(\hspace{7.91pt}-\hspace{7.91pt}) &- &\hspace{7.91pt}-\hspace{7.91pt}(\hspace{7.91pt}-\hspace{7.91pt}) &\hspace{7.91pt}-\hspace{7.91pt}(\hspace{7.91pt}-\hspace{7.91pt}) &- &0.59(0.13) &0.12(0.25) &0.17
\tabularnewline
\cline{1-10}

\end{tabular}
}
\caption{$\sigma$-distance values for all tested combinations of cosmological models and priors. This table compares results across the $\Lambda$CDM, $w$CDM and $w_0 w_a$CDM cosmological models, using both the standard prior and HOD-informed priors (TNHOD and UHOD). The numbers outside of the brackets represent results from the M+Q case, while the numbers inside the brackets correspond to the M+Q+H case. For the UHOD prior, only the M+Q case is shown.}
\label{tab:results}
\end{table}

\begin{figure}
    \centering
    \includegraphics[width=0.95\linewidth]{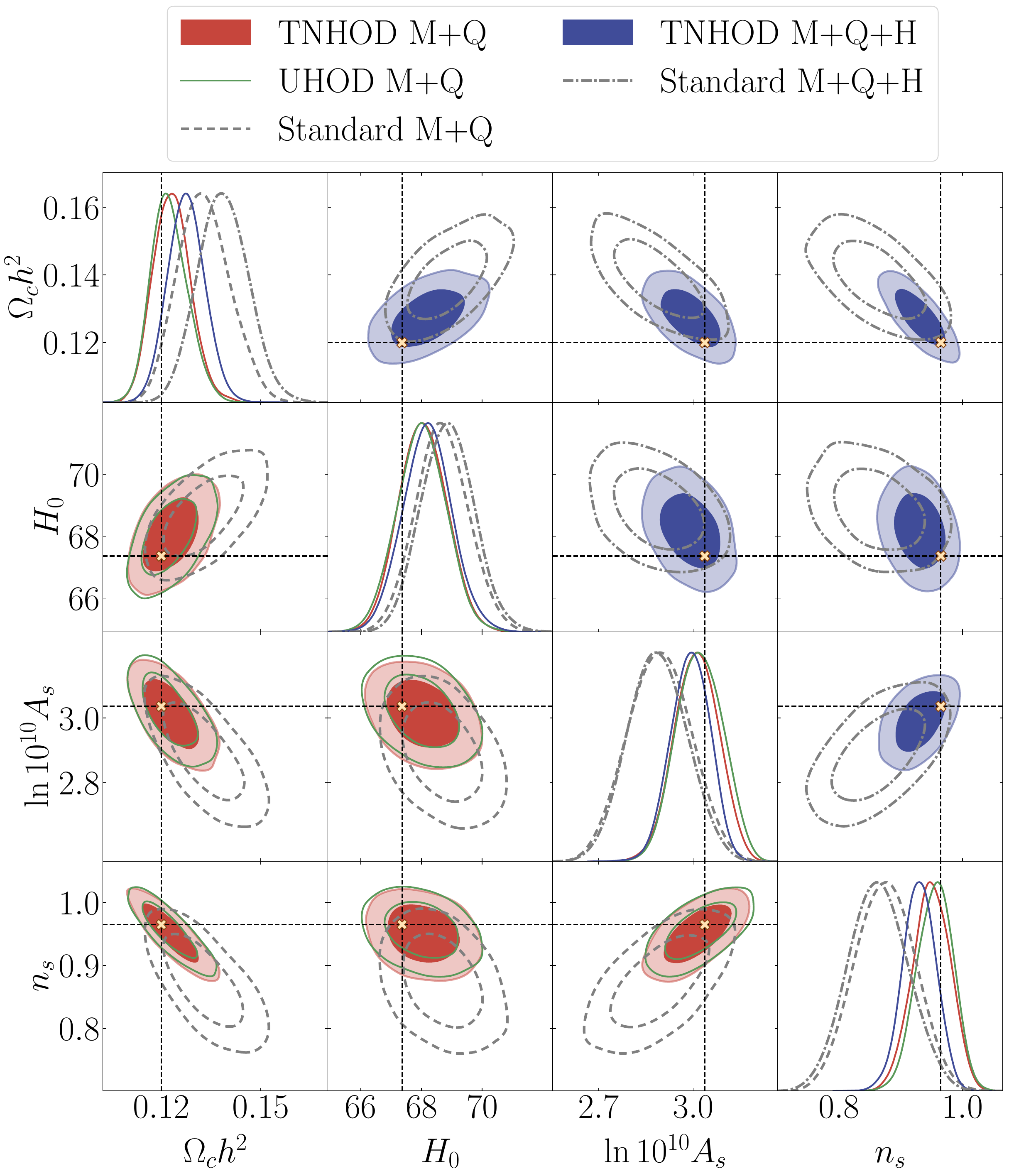}
    \caption{Marginalised posterior constraints obtained using different priors for $\Lambda \mathrm{CDM}$ model. The lower triangle shows the results using the monopole and quadrupole moments (M+Q), while the upper triangle includes the hexadecapole moment (M+Q+H). Red and green contours represent posteriors obtained with TNHOD and UHOD priors, respectively, for the M+Q case, while blue contours show TNHOD posteriors for the M+Q+H case. Dashed and dot-dash gray lines correspond to the standard prior results for M+Q and M+Q+H, respectively. Moccasin cross markers indicate the true values. All contours show 68$\%$ and 95$\%$ credible intervals.}
    \label{fig:compare_lcdm}
\end{figure}

\begin{figure}
    \centering
    \includegraphics[width=0.95\linewidth]{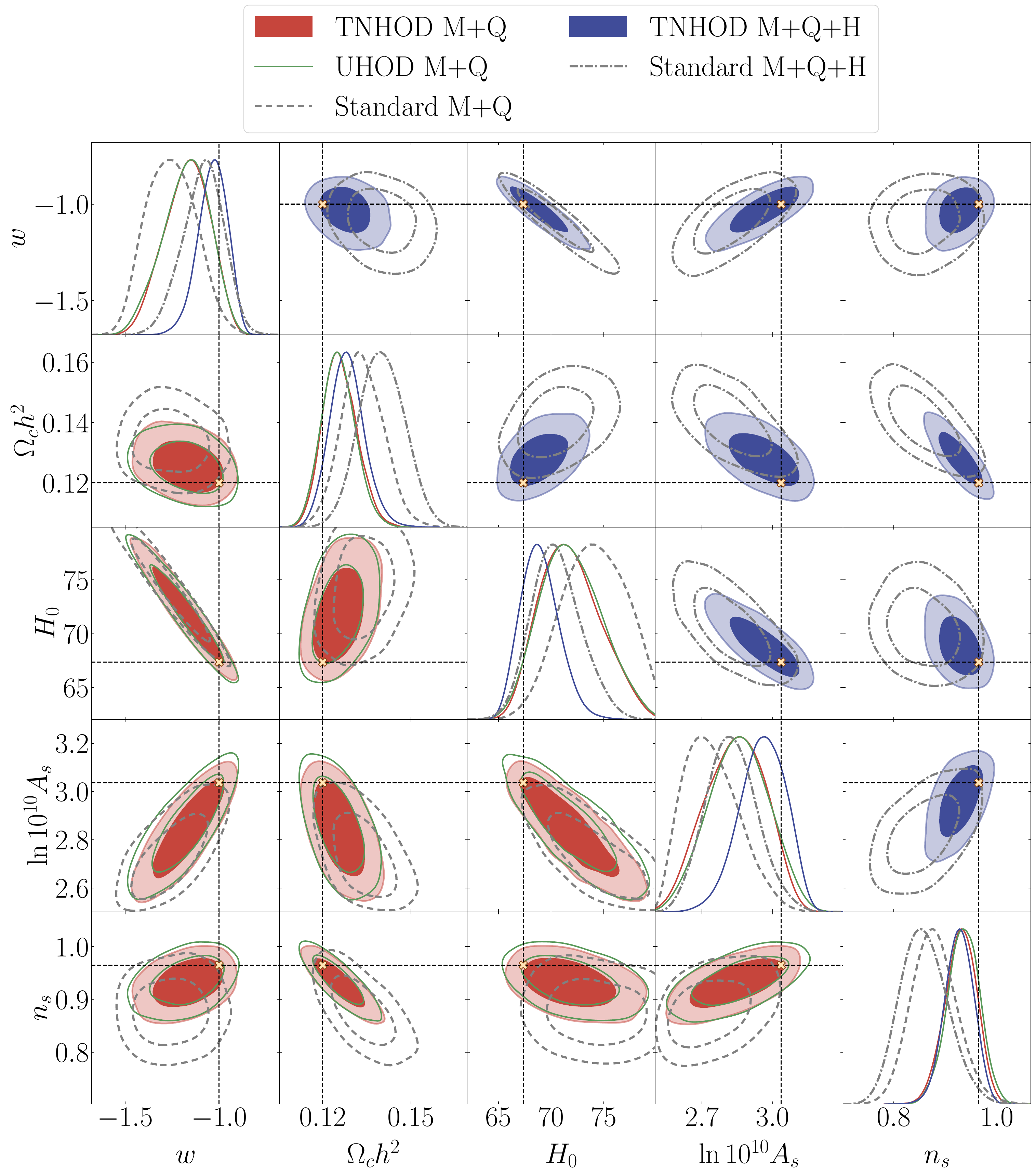}
    \caption{Similar plot as Figure~\ref{fig:compare_lcdm} for $w \mathrm{CDM}$}
    \label{fig:compare_wcdm}
\end{figure}

\begin{figure}
    \centering
    \includegraphics[width=0.95\linewidth]{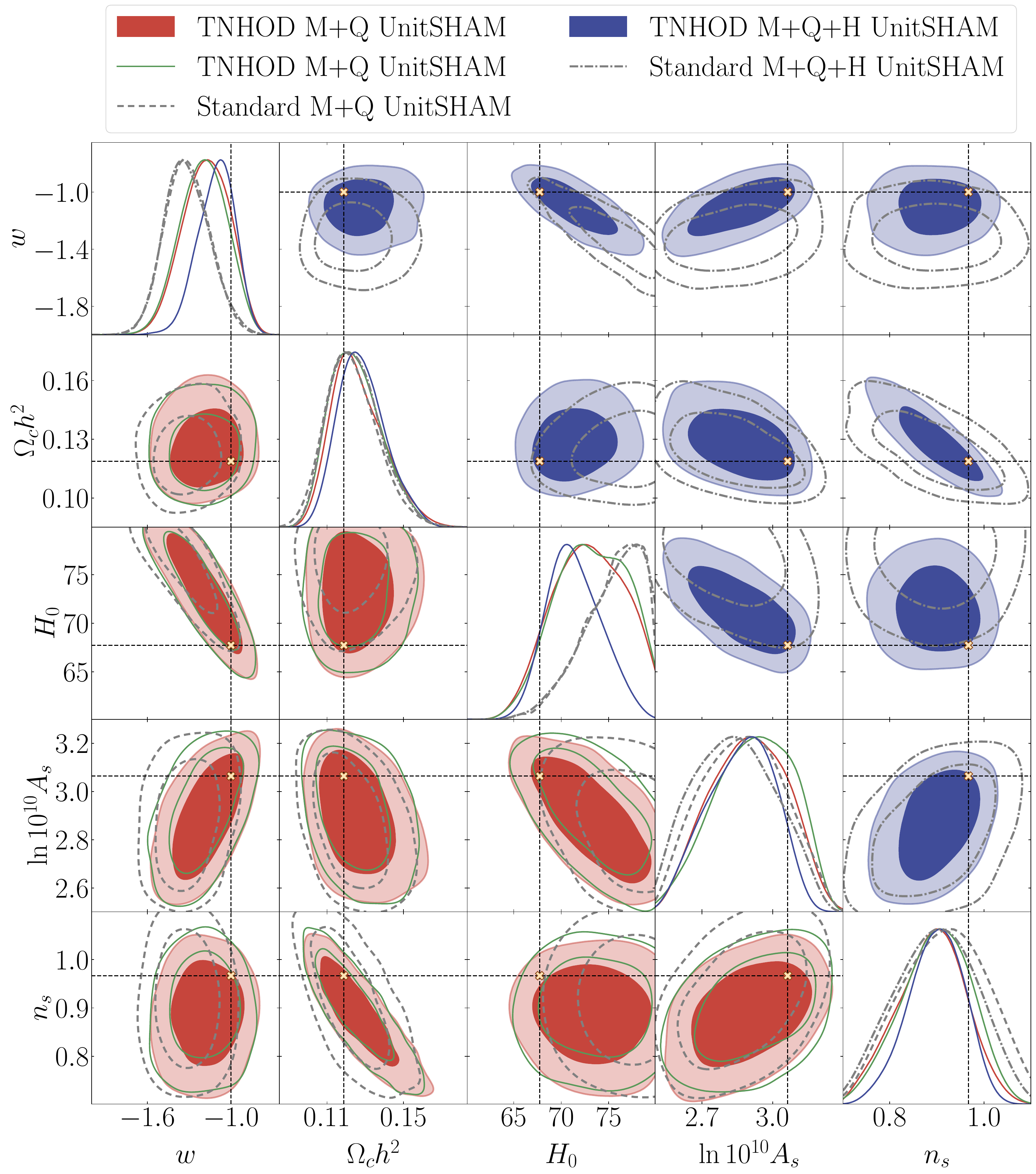}
    \caption{Similar plot as Figure~\ref{fig:compare_lcdm} for $w \mathrm{CDM}$ using a SHAM-Unit mock}
    \label{fig:compare_wcdm_unit}
\end{figure}

\begin{figure}
    \centering
    \includegraphics[width=0.95\linewidth]{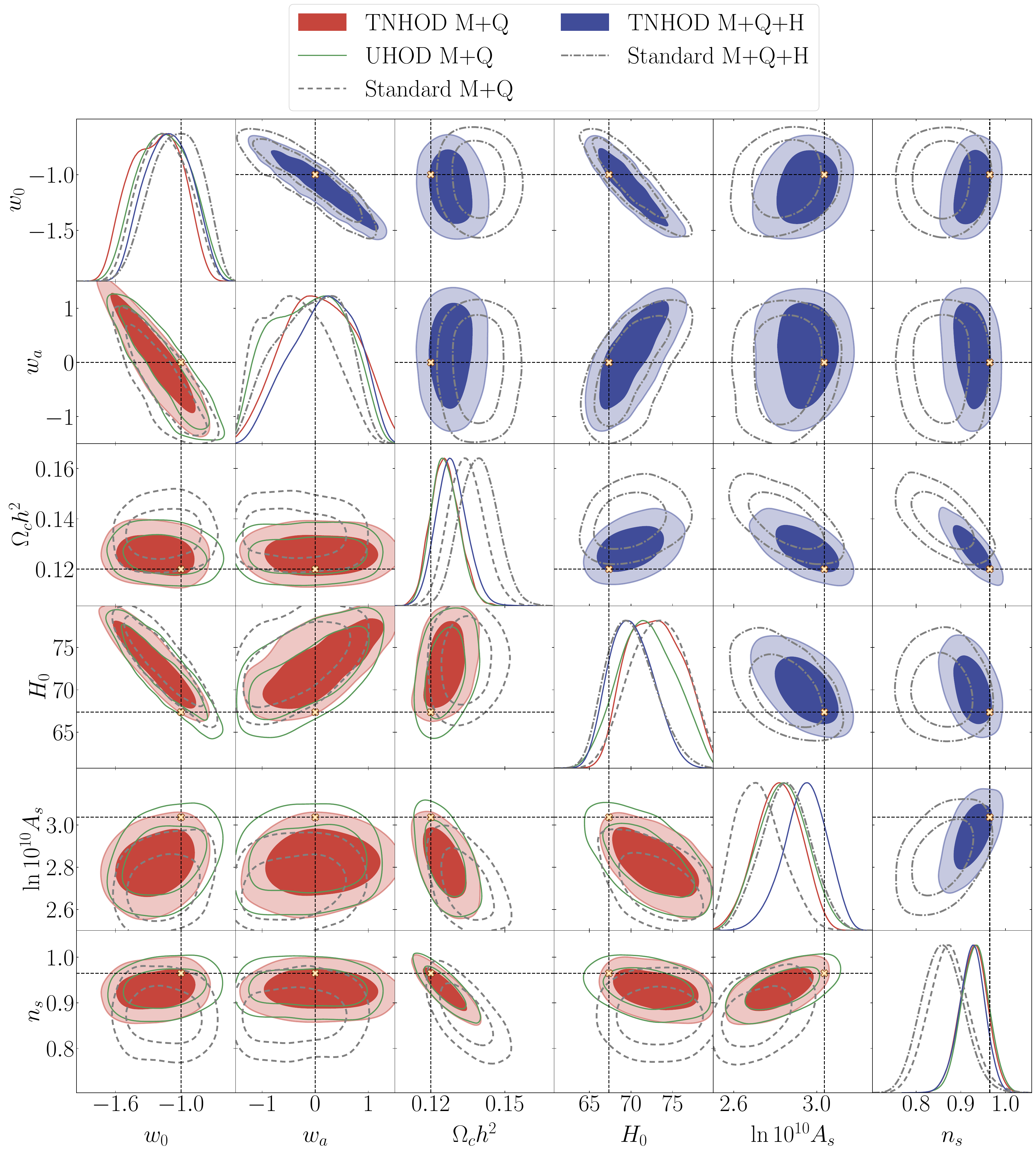}
    \caption{Similar plot as Figure~\ref{fig:compare_lcdm} for $w_0 w_a \mathrm{CDM}$}
    \label{fig:compare_w0wacdm}
\end{figure}

In this section, we perform a full-shape power spectrum analysis using the EFTofLSS redshift-space power spectrum model described in Section~\ref{subsec:eft} to evaluate the improvement of the HOD-informed prior compared to the standard prior in reducing prior effects. Both cosmological parameters and EFT parameters are allowed to vary, with a Big Bang Nucleosynthesis (BBN) prior \cite{Schoneberg2024TheUpdate} imposed on the physical baryon density $\Omega_b h^2$. The cosmological parameters sampled include: $\Omega_b h^2$, $\Omega_c h^2$, respectively physical baryon and cold dark matter densities, $H_0$, the Hubble constant, $\ln 10^{10} A_s$, the amplitude of primordial scalar fluctuations, and $n_s$, the scalar spectral index. 

To conduct this analysis, we prepared an DESI LRG-like HOD galaxy catalog using the \\ \texttt{AbacusSummit\char`_base\char`_c000\char`_ph000} box at redshift $z=0.8$. The HOD parameters used are based on those described in table~3 of \cite{Yuan2024TheABACUSSUMMIT} for LRG with $0.6<z<0.8$, with the galaxy density tuned to $5 \times 10^{-4}\,{\rm Mpc}^{-3}$. We computed the power spectrum multipoles for this catalog, adopting 18 linear bins with a scale cut of $0.02<k<0.2 \,h^{-1} \rm {Mpc}$ for the monopole, quadrupole, and hexadecapole moments. In the lower triangle of Figure~\ref{fig:hexa_hod_dp}, the cross markers indicate the underlying best-fit EFT parameter values for the mock power spectrum monopole and quadrupole, obtained with cosmological parameters fixed to their true values. As expected, most of these EFT parameters lie near the centers of the HOD-informed priors, which are specifically tailored to DESI-like LRGs. For the covariance matrix, we utilized the Gaussian Covariance from \textsc{CovaPT}, with linear bias, volume, and number density matching the catalog. 

We proceed to fit with three different cosmological models: $\Lambda \mathrm{CDM}$, $w\mathrm{CDM}$, $w_0 w_a \mathrm{CDM}$. For each model, we examine two cases: one using only the monopole and quadrupole moments (M+Q) and the other including the hexadecapole moment (M+Q+H). In the analyses without the hexadecapole, $c_\mathrm{r,2}$ and $c_\mathrm{\epsilon,1}$ are set to zero. The M+Q cases use the HOD-informed prior trained with M+Q data (red or green contours on Figure~\ref{fig:hexa_hod_dp}), while the M+Q+H cases use the HOD-informed prior trained with M+Q+H data (slate blue contours on Figure~\ref{fig:hexa_hod_dp}). We use the \textsc{pocoMC} as sampler\footnote{\url{https://github.com/minaskar/pocomc}} \cite{Karamanis2022AcceleratingCarlo,Karamanis2022PocoMC:Cosmology} and \textsc{GetDist}\footnote{\url{https://github.com/cmbant/getdist}} \cite{Lewis2019GetDist:Samples} to generate the contour plots. In this section we use \textsc{PyBird} to compute the theoretical model rather than \textsc{effort.jl}, to show that the emulator did not introduce any sizable error that could lead to biased results when using the HOD calibrated prior. We set the fiducial cosmology to match that of the simulation used to generate the mock. The Alcock-Paczynski effect is then applied by \textsc{PyBird} as described in Section \ref{subsec:ap}.

To quantitatively describe the discrepancy between the resulting posteriors using different priors and the true cosmology, we use the $\sigma$-distance metric. For a certain parameter $X$, $\sigma$-$\mathrm{distance}=\left|X_\mathrm{mean}-X_\mathrm{true}\right|/\sigma$, where $X_\mathrm{mean}$ is the marginalised mean, $X_\mathrm{true}$ is the true value from the simulation, and $\sigma$ is the 1-$\sigma$ error. While it is not necessary that the truth lies within a Bayesian credible interval a certain fraction of the time - it is a frequentist test - it can be indicative of a bias within the Bayesian analysis caused by the prior. Table~\ref{tab:results} shows the $\sigma$-distance values for all combinations of cosmological models and priors that have been tested. It is evident that the standard prior suffers significantly from prior effects, leading to a higher $\sigma$-distance and less accurate recovery of the true cosmology. In contrast, the HOD-informed priors (both TNHOD and UHOD) help mitigate these prior effects, consistently providing a more precise recovery of the true cosmological parameters across all models. We will discuss different cases in detail, along with the corresponding contour plots, in the following paragraphs.

Figure~\ref{fig:compare_lcdm} presents contour plots comparing the performance of HOD-informed priors against standard priors in a full-shape analysis for the $\Lambda \mathrm{CDM}$ cosmological model. The plot is divided into two sections: the lower triangle shows results from the analysis using monopole and quadrupole moments (M+Q), while the upper triangle includes the hexadecapole moment (M+Q+H). The red contours represent the TNHOD prior for the M+Q case, the green contours represent the UHOD prior for the M+Q case, and the blue contours correspond to the TNHOD prior for the M+Q+H case. The dashed and dot-dash gray contours represent the standard prior for the M+Q and M+Q+H cases, respectively. Cross markers indicate the true values of the cosmological parameters. Note that $\Omega_b h^2$ is dominated by the BBN prior, so it is omitted from the contour plots.

The results made using the standard prior are biased with the true cosmological parameters far from the credible regions. This is evident when comparing the HOD-informed prior contours (red, green, and blue) with the standard prior contours (gray dashed, gray dot-dashed) of Figure~\ref{fig:compare_lcdm}. 
On the other hand, all HOD-informed priors provide credible regions more consistent with the underlying cosmology, particularly for parameters such as $\Omega_c h^2$, $\ln 10^{10} A_s$ and $n_s$ compared to the standard prior. For example, focusing on the TNHOD M+Q (red) contours, the $\sigma$-distance for $\Omega_c h^2$ improves from 1.65 with the standard prior to 0.57. Similarly, for $\ln 10^{10} A_s$, the value improves from 1.46 to 0.37, and for $n_s$ , it drops from 1.88 to 0.50. Also for $H_0$, there is a notable improvement, with the $\sigma$-distance reduced from 1.55 to 0.87. These comparisons clearly demonstrate the ability of the HOD-informed prior to mitigate the prior effect. Note that though power spectrum alone has limited constraining power on $n_s$, it is still correlated with other parameters and EFT parameters. The HOD-informed prior, by reducing the degeneracies among those parameters, allows $n_s$ to be better recovered.
The HOD-informed prior leads to tighter constraints by $28\%$, $9\%$, $29\%$, and $37\%$ for $\Omega_c h^2$, $H_0$, $\ln 10^{10} A_s$, and $n_s$, respectively, compared to the standard prior. Thus we see that the $\sigma$-distance improvements were not driven by having larger credible intervals, but instead by mitigating prior effects. This improvement in constraints is not driven by physical knowledge of the particular halos being modelled, as we have used a wide range of HOD parameters.

Another interesting point is that the posteriors using TNHOD and UHOD are nearly identical, indicating that the cosmological analysis, at least for the $\Lambda$CDM model, is not significantly influenced by the distribution of HOD sample used to generate the HOD-informed prior. Although the UHOD prior exhibits a more concentrated distribution and the true $b_1$ lies beyond the 1-$\sigma$ interval of UHOD in the lower triangle of Figure~\ref{fig:hexa_hod_dp}, it does not produced biased or tighter constraints than the TNHOD prior. 

Figure~\ref{fig:compare_wcdm} present the contour plots for the $w\mathrm{CDM}$ model. In this case, the fit using the standard prior suffers from more severe prior effects due to the expanded parameter space. The $\sigma$-distance values for most parameters are around or exceed 2 in the M+Q case. Although the inclusion of the hexadecapole in the M+Q+H case provides some improvement, there are still modest
deviations, with $\Omega_c h^2$ and $n_s$ showing over 2-$\sigma$ differences from the true values. In contrast, the HOD-informed prior (red and blue contours) consistently recovers the true values across all parameters. Taking the TNHOD M+Q case as an example, the $\sigma$-distance for $\Omega_c h^2$ decreases from 1.90 with the standard prior to 1.03, for $H_0$ from 2.26 to 1.57, for $\ln 10^{10} A_s$ from 2.55 to 1.55, for $n_s$ from 1.97 to 1.13. and for the parameter $w$, from 1.97 to 1.42. The green (UHOD M+Q) and red (TNHOD M+Q) contours match each other perfectly, constraining power using HOD-informed prior remaining similar as using standard prior, and the arguments made for the $\Lambda$CDM case still hold for the $w$CDM model.

Figure~\ref{fig:compare_wcdm_unit} presents contour plots for the $w \mathrm{CDM}$ cosmological model, similar to Figure 5, but with the analysis performed on a galaxy catalog generated using the Sub-Halo Abundance Matching (SHAM) approach \cite{Kravtsov2004TheDistribution,Vale2004LinkingLuminosity,Conroy2006ModelingTime,Vale2006TheMass,Behroozi2010A4,Guo2016ModellingMatching}, also using a different simulation, UNIT \cite{Chuang2019UNITSurveys}. This SHAM-based mock catalog is an DESI LRG-like mock from \cite{Yu2024TheSimulation}, at redshift $z=0.8188$ with a box size of 1000$h^{-1} \mathrm{Mpc}$ per side. Even though the simulation is different, the results are consistent with those observed in Figure~\ref{fig:compare_wcdm}. We also fit this mock with $\Lambda \mathrm{CDM}$ and $w_0 w_a \mathrm{CDM}$ models, which both give similar results as using the Abacus mock. For brevity, we only show the contourplot for the $w \mathrm{CDM}$ model.
The fact that the HOD-informed prior, when applied to galaxy catalogs generated using different galaxy-halo connection approaches, gives consistent results, makes us more confident about applying it to real observational data.

The results from fitting the $w_0 w_a$CDM model are shown in Figure~\ref{fig:compare_w0wacdm}. Fits using the standard prior again lead to 1D credible intervals shifted from the truth, particularly for $\Omega_c h^2$, $\ln 10^{10} A_s$ and $n_s$. However, the inclusion of the hexadecapole seems to help in matching the credible intervals to the true values for $H_0$ and $\ln 10^{10} A_s$. Meanwhile, the HOD-informed priors continue to perform well, accurately matching intervals and the true cosmological parameters across the all cases. Interestingly, the UHOD prior (green contours) performs slightly better than the TNHOD prior (red contours) in this case, as indicated by the small shift in the 1 and 2 $\sigma$ contours. This is likely due to the use of a more flexible cosmological model and the fact that the data alone do not have enough constraining power. As a result, the more concentrated UHOD prior helped with a better recovery. However, the difference between the two is still small, and it is expected that when we combine more data from multiple redshifts, the dependency on the choice of the HOD sample distribution will again become minimal.

\section{Conclusion}
\label{sec:conclusion}

In this paper, we introduced an HOD-informed prior for the EFTofLSS framework to mitigate the prior effects commonly encountered in full-shape cosmological analyses of large-scale structure. We generated galaxy power spectra from a diverse set of HOD mock catalogs, covering a broad range of HOD parameter space and cosmological models. By fitting the EFTofLSS model to these power spectra, we extracted the best-fit EFT parameters and used their distribution to construct an HOD-informed prior. This approach provides a more physically motivated parameter space by mapping the EFT parameter space to a galaxy-halo connection framework. 

We tested the robustness of our HOD-informed prior by analyzing its dependence on cosmological models, HOD sample distributions, and the inclusion of the hexadecapole moment. Our results showed that the distribution of EFT parameters does not exhibit specific dependencies on cosmology, making the prior largely cosmology-independent. Importantly, the HOD sample distribution does influence the HOD-informed prior, as it intrinsically represents different types of tracers.

We then conducted a full-shape analysis on a DESI LRG-like HOD catalog generated at redshift $z=0.8$. In general, the HOD-informed prior consistently outperformed the standard prior, mitigating prior effects across all three cosmological models we tested and under various configurations. For example, in the $\Lambda$CDM case, the $\sigma$-distance for $\Omega_c h^2$ improved from 1.65 with the standard prior to 0.57 with the HOD-informed prior. Similarly, in the $w$CDM model, it decreased from 1.90 to 1.03. In the more flexible $w_0 w_a$CDM model, the HOD-informed prior brought the $\sigma$-distance down from 2.00 to 1.00. We achieved a good recovery of all cosmological parameters within credible regions, with detailed $\sigma$-distance values presented in Table~\ref{tab:results}. Those test on mock catalog highlight the robustness and effectiveness of the HOD-informed prior in recovering true cosmological parameters more precisely. 

We further tested this using a SHAM-based mock from the UNIT simulation, and the consistent results demonstrate that our approach is simulation-independent. We also find that, while the HOD sample distribution does influence the shape of the prior itself, it has limited impact on cosmological fitting. Together, these findings confirm the robustness of the HOD-informed prior and its suitability for application to real observational data.

In future work, we aim to extend the current HOD model by incorporating assembly bias, allowing us to capture additional nonlinear information. By accounting for these missing nonlinear modes that the baseline model overlooks, we can further refine the HOD-informed prior, making it even more reliable for cosmological analyses. While this study has focused on LRG-like galaxies, we plan to expand the method to other tracers by adjusting the HOD models and sample distributions, allowing our approach to be applied across different galaxy populations and redshifts. This method could be extended to higher-order statistics, such as the bispectrum for additional information. The 1-loop EFT model for the bispectrum has more nuisance parameters compared to power spectrum, making it more sensitive to prior effects \cite{DAmico:2022osl}.

As current-generation surveys continue to produce more data, and with the arrival of next-generation surveys, the advantages of full-shape analysis will become increasingly evident. It is crucial to extract accurate cosmological information from the available data to address key questions, such as the Hubble constant tension and the $S_8$ tension, from a large-scale structure perspective. The HOD-informed prior presented in this paper will play an important role in the future, making full-shape analysis performance better, providing a robust tool for both current and upcoming surveys.

\acknowledgments

We thank Niayesh Afshordi, Enzo Branchini, Pierre Burger, Pedro Carrilho, Shi-fan Chen, Héctor Gil Marín, Daniel Gruen, Jeffrey A. Newman, Ashley J. Ross, Uro\v{s} Seljak, Martin White, and Pierre Zhang for useful discussions and comments.

MB \& WP acknowledge the support of the Canadian Space Agency. WP also acknowledges support from the Natural Sciences and Engineering Research Council of Canada (NSERC), [funding reference number RGPIN-2019-03908].

Research at Perimeter Institute is supported in part by the Government of Canada through the Department of Innovation, Science and Economic Development Canada and by the Province of Ontario through the Ministry of Colleges and Universities.

This research was enabled in part by support provided by Compute Ontario (\url{https://www.computeontario.ca}) and the Digital Research Alliance of Canada (\url{https://alliancecan.ca/en}). 

We acknowledge the use of the NASA astrophysics data system \url{https://ui.adsabs.harvard.edu} and the arXiv open-access repository \url{https://arxiv.org}. The software was hosted on the GitHub platform \url{https://github.com}. The manuscript was typeset using the overleaf cloud-based LaTeX editor \url{https://www.overleaf.com}.


\bibliographystyle{JHEP}
\bibliography{hodeft_fixed}

\appendix
\section{Covariance matrix dependency}
\label{app:cov_dp}

\begin{figure}
    \centering
    \includegraphics[width=0.95\linewidth]{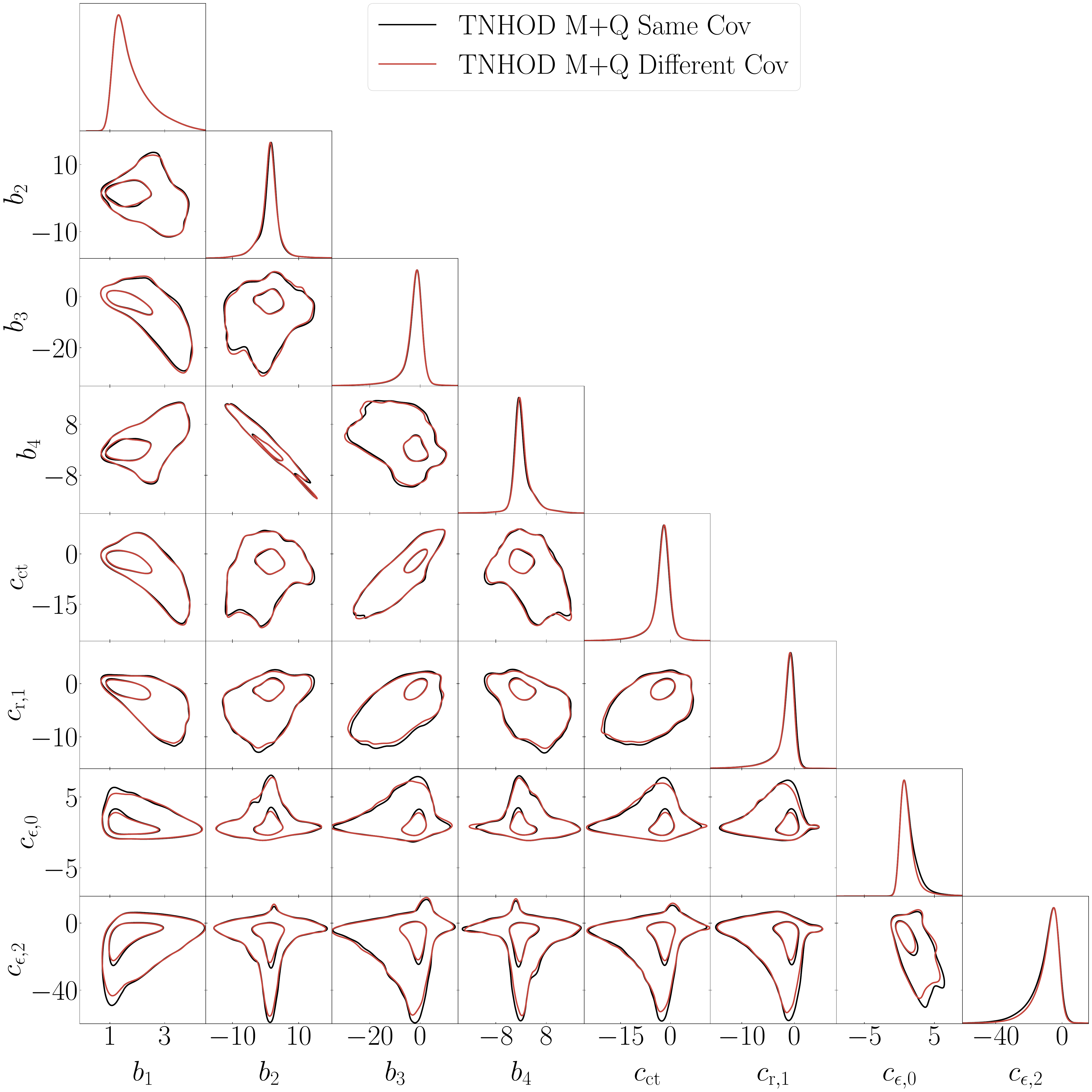}
    \caption{68$\%$ and 95$\%$ contour plots compare the best-fit EFT parameter distributions for the TNHOD M+Q case using a single covariance matrix (black contours) and using a different covariance matrix for each power spectrum (red contours). }
    \label{fig:covtest}
\end{figure}

We test here whether the best-fit EFT parameter distribution depends on the covariance matrix used during the fitting process. By default, we employ a single analytical Gaussian covariance matrix, assuming a linear bias of 2 and a galaxy density of $5 \times 10^{-4}\,{\rm Mpc}^{-3}$ for all fits. To investigate any dependence on the covariance matrix, we conduct a test for the TNHOD and M+Q case. 

For each power spectrum we fit, we generate an analytical Gaussian covariance matrix based on the true linear bias and number density of the corresponding HOD mock catalog. This results in using a different covariance matrix for every power spectrum, amounting to 320,000 different covariance matrices in total.

Figure~\ref{fig:covtest} shows that the best-fit EFT parameter distribution is nearly identical when using one single covariance matrix compared to using different covariance matrices for each fit. As our focus is on finding the best fits rather than exploring detailed EFT parameter distributions or degeneracies, it is safe to use the same covariance matrix for the best-fit finding. This also demonstrates that our EFT parameter best-fit distribution is effectively independent of the covariance matrix used.

\section{Check on the perturbativity of the prior}
\label{app:pert_check}
\begin{figure}
    \centering
    \includegraphics[width=0.95\linewidth]{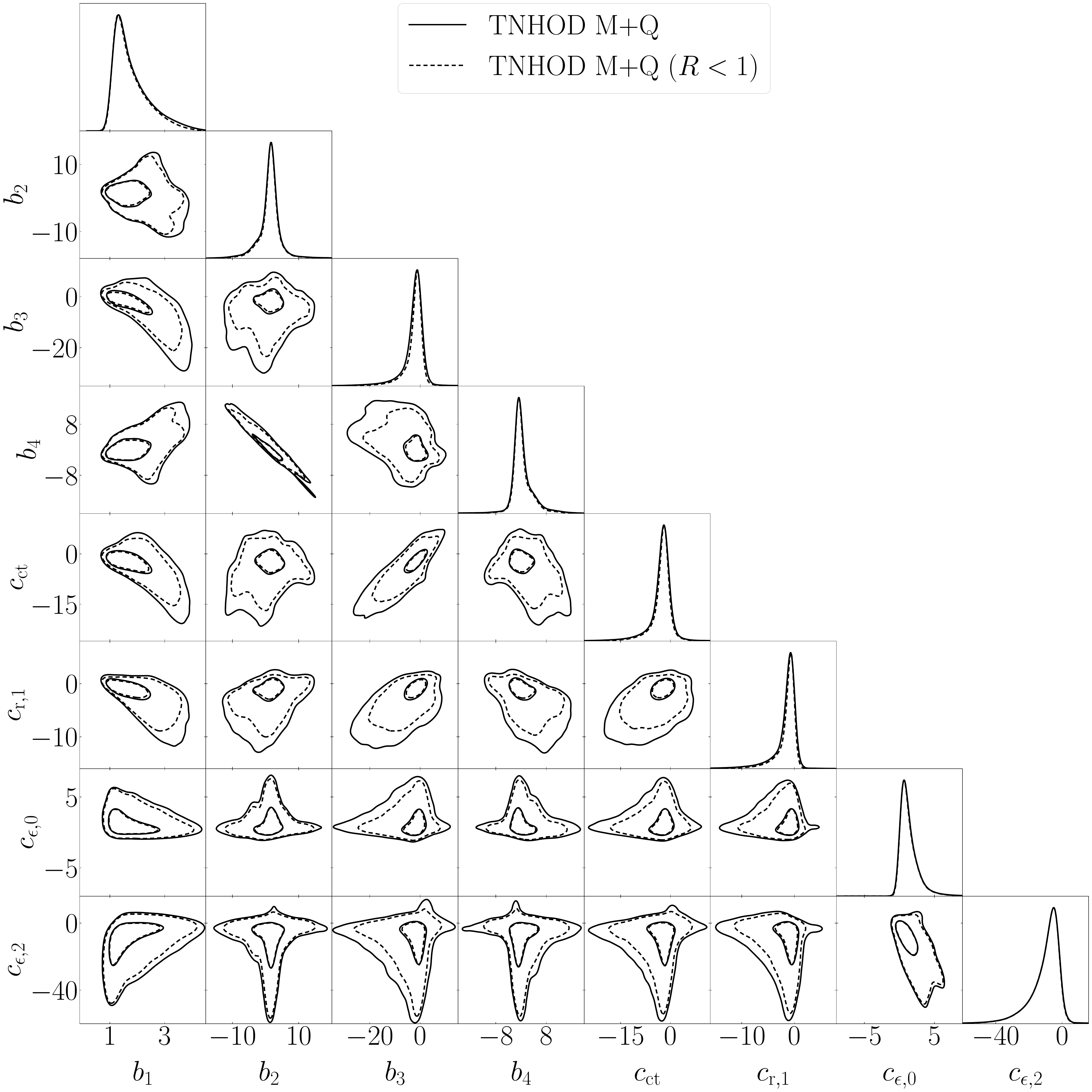}
    \caption{68$\%$ and 95$\%$ contour plots compare the best-fit EFT parameter distributions for the TNHOD M+Q case with (dashed contours) and without (solid contours) a perturbative cut.}
    \label{fig:pertchk}
\end{figure}

The EFTofLSS is based on the assumption that the model is being evaluated in the perturbative regime: if the contributions from the re-normalized loop-term happened to be bigger than the linear term, this would highlight a breakdown of the model and hence the non-reliability of the results.

In this appendix, we present the result of this check for the TNHOD M+Q case. For each best-fit combination, we compute the ratio 
\begin{equation}
    R=\frac{P_\text{1-loop}+P_\text{ct}}{P_\text{lin}},
\end{equation}
where the subscript \texttt{1-loop}, \texttt{ct}, and \texttt{lin} represent the 1-loop, counterterms, and linear contributions to the theoretical power spectrum monopole, respectively, at the highest $k$ considered in our fits. Only $3.1\%$ of the best fits show a ratio greater than 1, indicating that the vast majority of the best fits remain within the perturbative regime. We also would like to point out this be a consequence of the removal of best fits described in Sec.~\ref{subsec:nf}, when we removed points close to the boundary set on EFT parameters; given their large values, it was very likely they would have not passed the perturbativity test. Figure~\ref{fig:pertchk} shows the distributions of the best fits with (dashed) and without (solid) the perturbative cut. These two distributions exhibit nearly identical 1D marginalized distribution and 1-$\sigma$ contours, with only minor differences in their tails, as indicated by the difference in the 2-$\sigma$ contours. Given that the most of best-fit combinations comply with perturbative requirements and show minimal impact on the distribution, we conclude that using a prior with or without the perturbative cut does not significantly influence the analysis presented in this paper. 

\end{document}